\documentclass[a4paper, 12pt, titlepage]{article}

\usepackage{graphicx}
\usepackage{amsmath}
\usepackage{placeins}
\usepackage[colorlinks]{hyperref}
\usepackage{enumitem}
\usepackage[top=1cm, left=1.5cm, right = 1.5cm, bottom = 1.5cm]{geometry}
\begin{document}
	\title{Parametric and inverse analysis of flow inside an obstructed
		channel under the influence of magnetic field using physics
		informed neural networks}
		\author{E. Ghaderi\footnote{Department of Mechanical Engineering, Sharif University of Technology, Tehran, Iran}
		, M. Bijarchi\textsuperscript{*}\footnote{Assistant Professor, Department of Mechanical Engineering, Sharif University of Technology, Tehran, Iran. \textsuperscript{*}Responsible Author}
		, S. Kazemzadeh Hannani\footnote{Professor of Mechanical Engineering, Department of Mechanical Engineering Sharif University of Technology, Tehran Iran}
		, A. Nouri-Borujerdi\footnote{Professor of Mechanical Engineering, Department of Mechanical Engineering, Sharif University of Technology, Tehran Iran}}
		\maketitle
	\newpage
	\section*{Abstract}
	In this study, fluid flow inside of an obstructed channel under the influence of magnetic field has been analyzed using physics informed neural networks(PINNs). Governing equations have been utilized in low-order form and the solution has been obtained in dimensionless form. Geometric and physics-related dimensionless parameters have been used as input variables of the neural network in the learning process. The radius and longitudinal position of the obstruction have also been involved in the learning process and the problem has been solved parametrically. In the successive sections of the study, inverse problem has been a matter of interest, particularly in form of obtaining the Hartmann number using the proposed method. The results have indicated that the employed method determined the hartmann number with great accuracy and entailed proper results. In this study a thorough exploration of effects of physical and geometric parameters on the magnetically influenced flow through a duct has been carried out. The accuracy of the results obtained from the solving method proposed have been compared to results generated by common methods used in computational fluid mechanics. The parametric solution of the problem can serve as a powerful tool in optimization problems. The method was also applicable to cases where the parameters were outside of defined range, indicating it's generalization capabilities. \\
	\textbf{Keywords:} Physics Informed Neural Networks, Inverse problems, Hartmann number, Computational Fluid Mechanics
	\section{Introduction}
	Magnetohydrodynamics is the study of motion of electrically conductive fluids under the influence of external magnetic fields, which include liquid metals, electrolyte solution and ionized gases(Plasma)\cite{1}. Research carried out on combination of an electrically conductive flow inside an obstructed channels and electromagnetism has yielded impressive and inquisitive results. Heat transfer enhancement using magnetic nanofluids\cite{2} and prediction and treatment of stenosis.\cite{3} are examples of applications within this field. Researchers have used numerical methods such as FDM(Finite Difference Method), FEM(Finite Element Method) and FVM(Finite Volume Method) before, yet these methods come with several disadvantages such as complexity of the meshing process, increased computational costs required for problems having high-dimensional space, and the inherent errors caused by discretization. Conventional methods of numeric simulation are specifically hard to implement on some problems such as inverse problems and ill-posed problems,  as well as problems where a sizable domain has to be discretized, or where the boundary conditions are not sufficiently specified or the properties of the material are unknown\cite{4}. Discretizarion is a particularly crucial aspect of CFD simulations, which can lead to incorrect results if done poorly. One of the recent solutions to these problems is the utilization of artificial intelligence (AI) in numerical simulations, particularly Physics Informed Neural Networks(PINNs).\\
	AI is one of the most interesting fields of science, used substantially in areas such as natural language processing, Computer vision, Recommender systems and self-driving cars, as well as problems previously thought to be impossible to solve. Machine learning is one of the main branches of this field, employing neural networks with varying architectures such as MLP (Multilayer Perceptron), CNN(Convolutional Neural Network)
	and RNN(Artificial Neural Network)\cite{6}. In traditional methods employing use of AI, the model is trained using substantial amounts of data acquired by various means, hence they are called Data-driven methods\cite{7}. Unfortunately, in problems concerning fluid mechanics and heat transfer little information is available since acquiring the data requires complex simulations or sophisticated experiments. Moreover, in these models the influence of the physical laws and problem premises are ignored\cite{7}, while most models concerning fluid mechanics and heat transfer are described by conservation laws and differential equations, making them important to researchers\cite{4}. These challenges lead to developing a new branch in the field, namely the Scientific Machine Learning.\cite{8}\\
	PINN and practical use of machine learning in this field were introduced by Raissi et al for the first time.\cite{9} In this method, differential equations and their initial and boundary conditions are imposed as physical constraints, guiding the learning process and making it's predictions more accurate, even when no results are available. Raissi et al\cite{10} have previously studied governing equations and explored it's use in obtaining pressure distributions and velocity fields for cases where data was difficult to acquire, such as cases involving aerodynamics of a plane or blood vessels. In other studies, use of PINN for cases such as flow around an immersed cylinder\cite{11}, boundary layer problems\cite{12}, modeling turbulence\cite{13} and battery temperature estimation\cite{14} has been explored. Additionally, several studies have been carried out to further improve and accelerate these methods.[17-15]. These studies have yielded interesting results and entertained the possibility of using this method as a powerful tool for analyzing heat transfer and fluid dynamics problems.\\
	Reviewing the literature of this topic reveals that due to novelty of this method and inherent complexity of flow in obstructed channels under the influence of magnetic fields, no proper research has been done on this specific problem. In this research, parametric analysis of obstructed channels under the influence of magnetic fields has been carried out using PINN‌ and influence of dimensionless parameters such as  Reynolds number, Hartmann number and dimensionless groups regarding geometric specifications such as placement and radius of obstruction has been studied. Ability of the presented model to yield results outside of specified limits is one of the novelties of the article at hand. By using the dimensionless and low order form of the equations, speed and precision of the results is enhanced. Hartmann number has been calculated as a parameter of interest in the section covering the inverse problem.
	\section{Problem description}
\begin{figure}[h]
		\centering
		\includegraphics[scale=1]{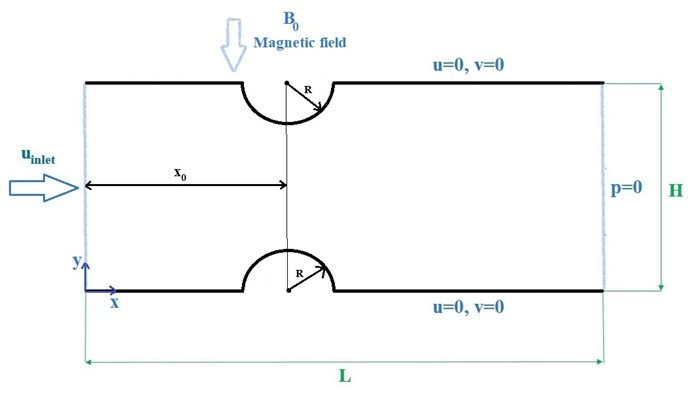}
		\caption{Geometric specifications and boundary conditions}
		\label{Fig1}
	\end{figure}

	In this study, a magnetically influenced flow passing through an obstructed channel is analyzed. Geometry of the fluid domain and boundary conditions are described in Fig.\ref{Fig1}. In this figure, $p$, $u$ and $v$ are pressure and velocity components in x and y direction respectively. $H$, $L$, $R$ and $x_0$ also represent channel width, length, obstruction radius and its placement. As the fluid enters the channel, characteristics of the flow changes according to the obstruction parameters and magnetic field.  
	\FloatBarrier
	As the continuum mechanics principles are applicable to this problem, governing differential equations for the flow can be described using equations \ref{Eq1} to \ref{Eq3}.\cite{1} In order to derive these equations, fluid is assumed to be homogeneous, single phase and having constant properties throughout the domain.
	 The flow is modeled as Newtonian, incompressible and laminar. A steady-state 2-D analysis is performed on this problem, neglecting gravity. External electric fields and inducted magnetic fields are also neglected.\cite{18} Magnetic and electric fields effect each other. A magnetic field in a electrically conductive medium induces an electric field, which conceives a new magnetic field [19]. However, if the magnetic Reynolds number is small, these effects of electromagnetic inductance pale in comparison to the externally applied magnetic field
	\begin{equation}
		\frac{\partial u}{\partial x}+\frac{\partial v}{\partial y}=0
		\label{Eq1}
	\end{equation}
	\begin{equation}
	u\frac{\partial\ u}{\partial\ x}+v\frac{\partial\ u}{\partial\ y}=-\frac{1}{\rho}\frac{\partial\ p}{\partial\ x}+\upsilon\left(\frac{\partial^2u}{\partial\ x^2}+\frac{\partial^2u}{\partial\ y^2}\right)-\frac{\sigma\ B_0^2}{\rho}u
	\end{equation}
	\begin{equation}
	u\frac{\partial\ v}{\partial\ x}+v\frac{\partial\ v}{\partial\ y}=-\frac{1}{\rho}\frac{\partial\ p}{\partial\ y}+\upsilon\left(\frac{\partial^2v}{\partial\ x^2}+\frac{\partial^2v}{\partial\ y^2}\right)
		\label{Eq3}
	\end{equation}
	In the above equations, $\rho$, $\nu$, $B_0$ and $\sigma$ represent density, viscosity, external uniform magnetic field and electric conductivity of the fluid respectively. 
	\section{Solving method}
	In this section, the described governing equations and boundary conditions are analyzed
	using the PINN method.
	\subsection{Governing equations}
	In studies conducted by Rao et al.\cite{11}, Laubscher et al.\cite{21} and Hu et al.\cite{22}, using lower order differential equations is recommended when using PINN. Thus, In order to reduce the computational cost of the backward differentiation and better implement the Neumann boundary conditions if necessary, governing equations have been reformulated in the form of first-order derivatives.(Eq.\ref{Eq4}-Eq.\ref{Eq6}). Furthermoer, if the properties of the fluid are not constant, such as cases where the viscosity changes with temperature or non-Newtonian fluids, the proposed method can be applied effectively.
	\begin{equation}
		\frac{\partial u}{\partial x}+\frac{\partial v}{\partial y}=0
		\label{Eq4}
	\end{equation}
	\begin{equation}
		u\frac{\partial u}{\partial x}+v\frac{\partial u}{\partial y}=\frac{1}{\rho}(\frac{\partial \tau_{11}}{\partial x}+\frac{\partial \tau_{12}}{\partial y})-\frac{\sigma B_0^2 u}{\rho}
		\label{Eq5}
	\end{equation}
	\begin{equation}
		u\frac{\partial v}{\partial x}+v\frac{\partial v}{\partial y}=\frac{1}{\rho}(\frac{\partial \tau_{12}}{\partial x}+\frac{\tau_{22}}{\partial y})
		\label{Eq6}
	\end{equation}
	In the above equations, $\tau$ is defined as the 2-D Cauchy stress tensor defined as follows:
	\begin{equation}
		\tau_{11}=-p+2\rho \nu\frac{\partial u}{\partial x}
		\label{7}
	\end{equation}
	\begin{equation}
		\tau_{22}=-p+2\rho \nu\frac{\partial v}{\partial y}
		\label{8}
	\end{equation}
	\begin{equation}
		\tau_{12}=\rho \nu (\frac{\partial u}{\partial y}+\frac{\partial v}{\partial x})
	\end{equation}	
	to convert the preceding equations to dimensionless form, the following groups are defined:
	\begin{equation}
	\begin{split}
		&x^ \ast = \frac{x}{H},y^ \ast=\frac{y}{H},u^ \ast=\frac{u}{u_{max}},v^ \ast=\frac{v}{v_{max}},p^ \ast=\frac{p}{\rho u_{max}^2}\\
		&\tau_{11}^ \ast=\frac{\tau_{11}}{\rho u_{max}^2},\tau_{12}^ \ast=\frac{\tau_{12}}{\rho u_{max}^2},\tau_{22}^ \ast=\frac{\tau_{22}}{\rho u_{max}^2}
	\end{split}
	\end{equation}
	where $u_{max}$ is defined as the maximum velocity at the center of the parabolic velocity profile, namely the inlet velocity profile in uniform flow.
	The governing equations can now be written in dimensionless form using the above dimensionless groups, which form famous dimensionless groups such as Reynolds number($Re=\frac{U_{max}H}{\nu}$) and Hartmann number($Ha=B_0H\sqrt{\dfrac{\sigma}{\rho \vartheta}}$):
		\begin{equation}
		\frac{\partial u^ \ast}{\partial x^ \ast}+\frac{\partial v^ \ast}{\partial y^ \ast}=0
		\label{Eq11}
	\end{equation}
	\begin{equation}
		u^ \ast\frac{\partial u^ \ast}{\partial x^ \ast}+v^ \ast\frac{\partial u^ \ast}{\partial y^ \ast}=(\frac{\partial \tau_{11}^ \ast}{\partial x^ \ast}+\frac{\partial \tau_{12}^ \ast}{\partial y^ \ast})-\frac{Ha^2}{Re}u^ \ast
		\label{Eq12}
	\end{equation}
	\begin{equation}
		u^ \ast\frac{\partial v^ \ast}{\partial x^ \ast}+v^ \ast\frac{\partial v^ \ast}{\partial y^ \ast}=(\frac{\partial \tau_{12}^ \ast}{\partial x^ \ast}+\frac{\tau_{22}^ \ast}{\partial y^ \ast})
		\label{Eq13}
	\end{equation}
		\begin{equation}
		\tau_{11}^ \ast=-p^ \ast+2/Re\frac{\partial u^\ast}{\partial x^\ast}
		\label{14}
	\end{equation}
	\begin{equation}
		\tau_{22}^ \ast=-p^ \ast+2/Re\frac{\partial v^ \ast}{\partial y^ \ast}
	\end{equation}
	\begin{equation}
		\tau_{12}^ \ast=\frac{1}{Re}(\frac{\partial u^ \ast}{\partial y^ \ast}+\frac{\partial v^ \ast}{\partial x^ \ast})
	\end{equation}
	\subsection{PINN}
\begin{figure}[h]
		\centering
		\includegraphics[scale=1]{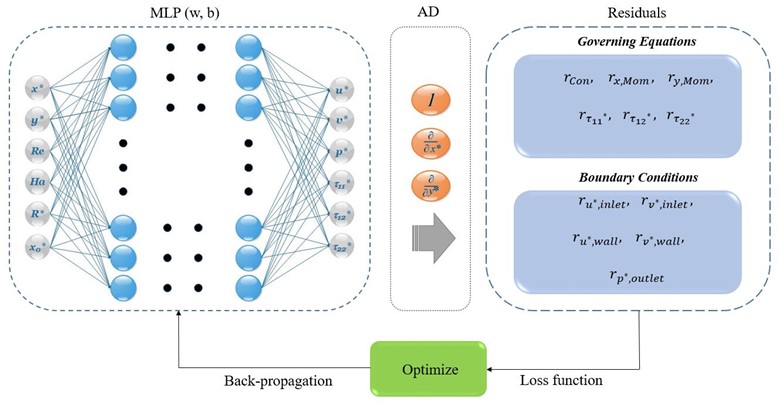}
		\caption{Schematics of the PINN‌ used in this study}
		\label{Fig2}
	\end{figure}

\begin{figure}[h]
		\centering
		\includegraphics[scale=1]{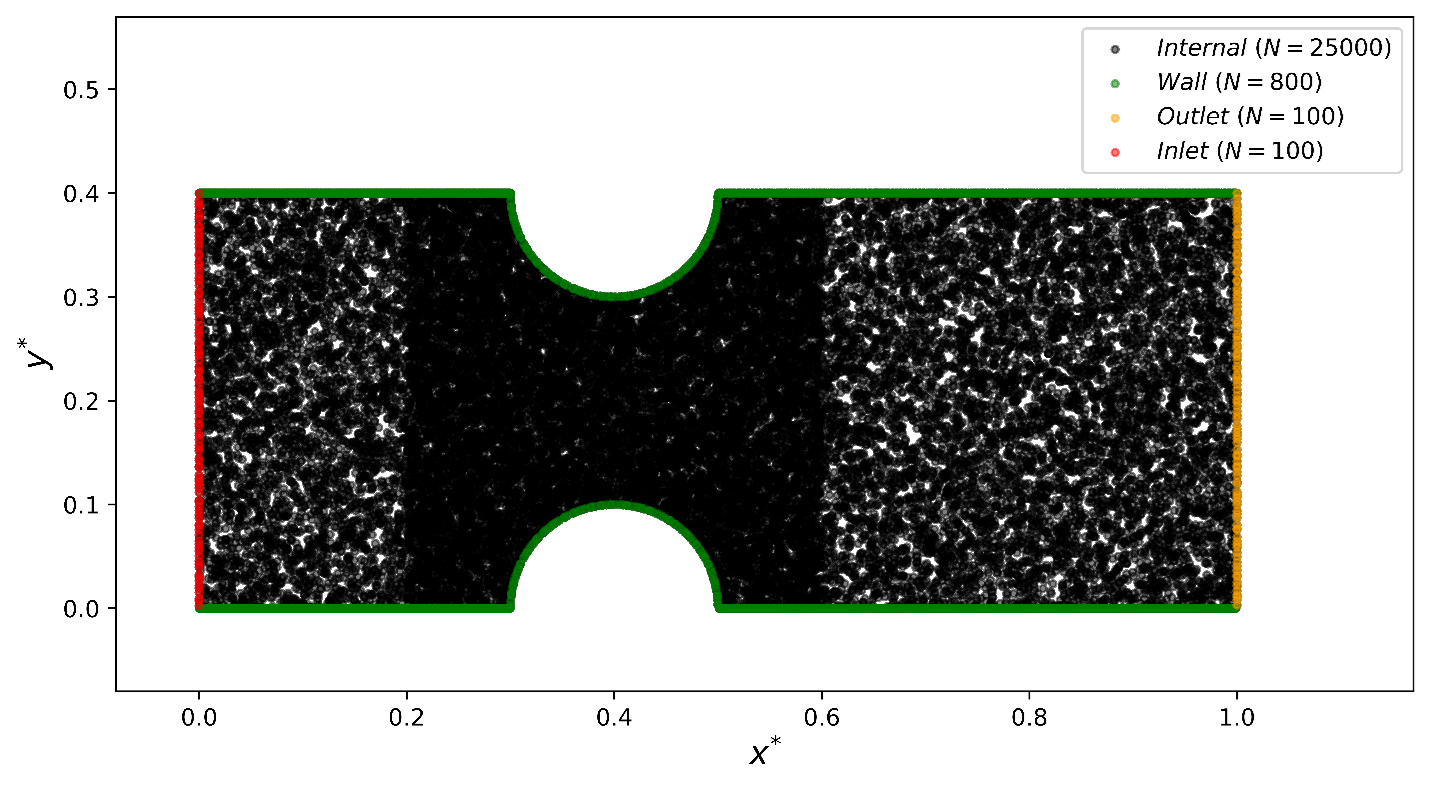}
		\caption{Points generated using LHS method for use in governing equations and boundary conditions}
		\label{Fig3}
	\end{figure}

	As stated in the previous section, in order to simulate the fluid flow in the obstructed channel under the influence of magnetic fields, partial differential equations have been used. According to the introduction where the weaknesses of conventional computational methods were pointed out, PINN‌ was used in this study. In this method, the loss function of the neural network is derived from the boundary condition and physics of the phenomenon. As depicted in Fig.\ref{Fig2}, the input layer consists of positional($x^ \ast,y^ \ast$), geometric($x_0^ \ast,R^ \ast$) and physical dimensionless parameters(Re, Ha) , while the output layer includes dimensionless parameters related to the velocity fields($u^\ast,v^\ast$), pressure distribution($p^\ast$) and Cauchy stress tensor parameters($\tau_{22}^ \ast,\tau_{12}^ \ast,\tau_{11}^ \ast$). Required Derivatives used in the equations are computed using the Automatic Derivative method(AD)\cite{23}. Also, in order to incorporate the nonlinearity and complexities of this flow influenced by magnetic fields in the model, the hyperbolic tangent(tanh) activation function has been applied as a nonlinear activation function to each neuron in the network.
	\FloatBarrier
	\FloatBarrier
	In subsequent steps, the loss function is formulated by utilizing the governing equations of the physics problem and the boundary conditions. In this method, instead of discretization and refining the grid, random points from the problem domain and boundaries are selected the loss associated with each of these points should be minimized. In this study, these random points have been chosen using LHS(Latin Hypercube Sampling), as depicted in Fig.\ref{Fig3}.\cite{24} It is noteworthy that in order to improve the precision of the results, more sampling points were selected from the parts of the domain closer to the obstruction. In the 2-D simulation, 25000 points were chosen in the computational domain and 1000 points were placed on the boundaries. The position of these points in shown in Fig.\ref{Fig3}. Additionally, the problem was analyzed in a 6-dimensional form, necessitating the consideration of a higher number of points.
	\FloatBarrier
	\FloatBarrier
	At the start of the solving procedure, the weights of the neural network were initialized using Xavier initialization.\cite{25} Updating the network parameters including weights and biases is carried out by optimization algorithms such as ADAM\cite{26} and L\_BFGS\cite{27}. The computations continue until the value of the total loss function is less than the desired error value.\\
	In order to create the loss function in this study, first the residuals are calculated using Eq\ref{Eq17} to Eq\ref{Eq27}. Equation \ref{Eq17} to \ref{Eq22} are related to the governing equations, while equations \ref{Eq23} to \ref{Eq27} are related to the boundary conditions of the problem.
	
	\begin{align}
		&r_{con}=\frac{\partial u^ \ast}{\partial x^ \ast}+\frac{\partial u^ \ast}{\partial y^ \ast}
		\label{Eq17}\\
		&r_{xmom}=u^ \ast\frac{\partial u^ \ast}{\partial x^ \ast}+v^ \ast\frac{\partial u^ \ast}{\partial y^ \ast}-(\frac{\partial \tau_{11}^ \ast}{\partial x^ \ast}+\frac{\partial \tau_{12}^ \ast}{\partial y^ \ast})+\frac{Ha^2}{Re}u^ \ast\\
		&r_{y,Mom}=u^\ast\frac{\partial v^\ast}{\partial x^\ast}+v^\ast\frac{\partial v^\ast}{\partial y^\ast}-\left(\frac{\partial{\tau_{12}}^\ast}{\partial x^\ast}+\frac{\partial{\tau_{22}}^\ast}{\partial y^\ast}\right)\\
		&r_{{\tau_{11}}^\ast}=-p^\ast+\frac{2}{Re}\frac{\partial u^\ast}{\partial x^\ast}-{\tau_{11}}^\ast\\
		&r_{{\tau_{12}}^\ast}=-p^\ast+\frac{2}{Re}\frac{\partial v^\ast}{\partial y^\ast}-{\tau_{22}}^\ast\\
		\label{Eq22}
		&r_{{\tau_{22}}^\ast}=\frac{1}{Re}(\frac{\partial u^\ast}{\partial y^\ast}+\frac{\partial v^\ast}{\partial x^\ast})-{\tau_{12}}^\ast\\
		\label{Eq23}
		&r_{u^\ast,inlet}=u^\ast\mathrm{\mathrm{-}}\frac{u_{\mathrm{inlet}}}{\mathrm{u}_{\mathrm{max}}}\\
		&r_{v^\ast,inlet}=v^\ast\\
		&r_{u^\ast,wall}=u^\ast\\
		&r_{v^\ast,wall}=v^\ast\\
		&r_{p^\ast,outlet}=p^\ast\mathrm{\mathrm{-}}\frac{p_{\mathrm{outlet}}}{\mathrm{\rho}u_{max}^2}
		\label{Eq27}
	\end{align}
	In the equations above, r represents the value of the residuals. The loss function associated with governing equations, boundary condition and total loss function are defined below:
	\begin{equation}
			{Loss}_{PDE}=\frac{1}{N_{PDE}}\sum_{j=1}^{N_{PDE}}({r_{Con}}^2+{r_{x,Mom}}^2+{r_{y,Mom}}^2+{r_{{\tau_{11}}^\ast}}^2+{r_{{\tau_{12}}^\ast}}^2+{r_{{\tau_{22}}^\ast}}^2)
	\end{equation}
	\begin{equation} 
	\begin{split}
	&{Loss}_{BC}=\frac{1}{N_{BC,inlet}}\sum_{j=1}^{N_{BC,inlet}}({r_{u^\ast,inlet}}^2+{r_{v^\ast,inlet}}^2)+\\
		 &\frac{1}{N_{BC,wall}}\sum_{j=1}^{N_{BC,wall}}{({r_{u^\ast,wall}}^2+{r_{v^\ast,wall}}^2)+\frac{1}{N_{BC,outlet}}\sum_{j=1}^{N_{BC,outlet}}({r_{p^\ast,outlet}}^2)}
	\label{Eq29}
	\end{split}
	\end{equation}
	\begin{equation}
		{Loss}_{Total}={Loss}_{PDE}+{\beta Loss}_{BC}
	\end{equation}
	in the article at hand, $\beta$ coefficient has been assumed 1. The computations terminate when the total loss is less than the desired error value.
	\section{Results and discussion}
\begin{figure}[h]
		\centering
		\includegraphics[scale=1]{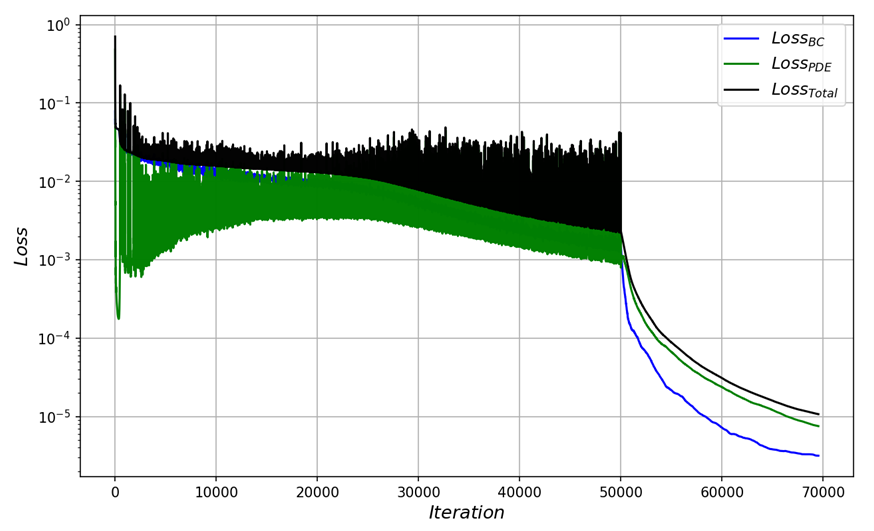}
		\caption{loss functions of the PINN‌ as the solution progressed}
		\label{Fig4}
	\end{figure}

	In this section the PINN‌ described in Fig.\ref{Fig2} is used to analyze the magnetically influenced flow in an obstructed channel. The algorithms were implemented in python\cite{28} using Pytorch library.\cite{29}  The total loss function including loss related to the boundary conditions and governing equations is depicted in Fig.\ref{Fig4}. Given that the parameters of the problem were dimensionless, weight of each term in the total loss function is assumed 1. Trend of the loss function indicates that the solution has converged with an error of $10^{-4}$. The sudden change perceived in the 50000th iteration is due to the shift from the ADAM‌ algorithm to L\_BFGS‌ algorithm. According to the recommendations disclosed by Rao et al.\cite{11}, Eiyazi et al.\cite{13} and Biswas and Anand\cite{30}, ADAM optimization algorithm was used in the initial iteration to avoid getting stuck in the local minima. As the solution progressed, L\_BFGS method was employed to achieve better convergence and greater precision. 
	\FloatBarrier
	\FloatBarrier
	\subsection{Independency from the number of layers and their respective number of neurons}
	In order to obtain a solution, some points in the domain are considered, where the governing equations have to be satisfied. Unlike conventional CFD methods, no discretization is needed, since these points are randomally selected using LHS‌ algorithm. Optimal placement and specification of points is a subject of interest in this article. In order to accurately model complex flows, more points need to be positioned near the regions where the changes are more prominent.\cite{13} Since a six dimensional model has been developed for this problem, several simulations have been carried out to assess the independency of the solution from points. Results obtained from these simulations are presented in Table \ref{Table1}
	\begin{table}[h]
		\centering
		\begin{tabular}{|c|c|c|c|}
			\hline
			PINN Loss function&execution time&Points on boundaries&points inside the domain\\
			\hline
			$9\times10^{-4}$&25min. 46s&500&12500\\
			\hline
			$2\times10^{-4}$&39min. 02s&1000&25000\\
			\hline
			$1\times10^{-4}$&54min. 10s&2000&50000\\
			\hline
			$1\times10^{-4}$&71min. 04s&3000&75000\\
			\hline
			$7\times10^{-4}$&112min. 43s&4000&100000\\
			\hline
		\end{tabular}
		\caption{Effect of different point distributions on using PINN‌ for this problem}
		\label{Table1}
	\end{table}
	in Table \ref{Table1} the effect of different number of points on the loss function and execution time is shown. As the number of points increases, the execution time increases, yet the loss function is not severely affected, since the algorithms can not decrease the loss function for all points effectively.\\
	In addition to sampling points, proper number of hidden layers and number of neurons per layer was a major topic of interest. In order to obtain the network with the simplest structure capable of proper prediction, a series of networks each varying in hidden layer arrangement and number of neurons per layer have been examined, ranging from only one hidden layer with ten neurons to nine layers each having 80 neurons. Errors regarding prediction of thermal field for these networks have been studied and it can be inferred that deeper networks(having more hidden layers and more neurons per layer) are more accurate, yet optimized number of layers and neurons per hidden layer has to be employed such that as the size of the network increases, the changes in the parameters become negligible\cite{21}. In Table\ref{Table2} the effect of the number of hidden layers and neurons per layer is shown, indicating that as the network becomes deeper, a nonlinear behavior is exhibited by the network, enabling it to obtain better solutions. Yet, after hitting an optimum the size of the neural network does not affect the accuracy of the solution much, but the execution time increases significantly. For instance, by increasing the number of hidden layers to 7 and neurons to 60, the solution takes much longer to converge, yet no significant decrease in loss function is perceived.  
	\begin{table}[h]
		\centering
		\begin{tabular}{|c|c|c|c|}
			\hline
			Loss function&execution time&Number of neurons&Number of hidden layers\\
			\hline
			$3\times10^{-4}$&18min. 51s&10&1\\
			\hline
			$5\times10^{-4}$&37min. 04s&20&3\\
			\hline
			$1\times10^{-4}$&54min. 10s&40&5\\
			\hline
			$9\times10^{-5}$&69min. 45s&60&7\\
			\hline
			$4\times10^{-5}$&97min. 02s&80&9\\
			\hline
		\end{tabular}
		\caption{number of hidden layers and number of neurons in each layer using PINN}
		\label{Table2}
	\end{table}
	According to the conducted tests, ideal number of points, neurons and layers has been set to 52000, 40 and 5 respectively.
	\subsection{Investigating dimensionless velocity components and pressure}
\begin{figure}[h]
		\centering
		\includegraphics[scale=1]{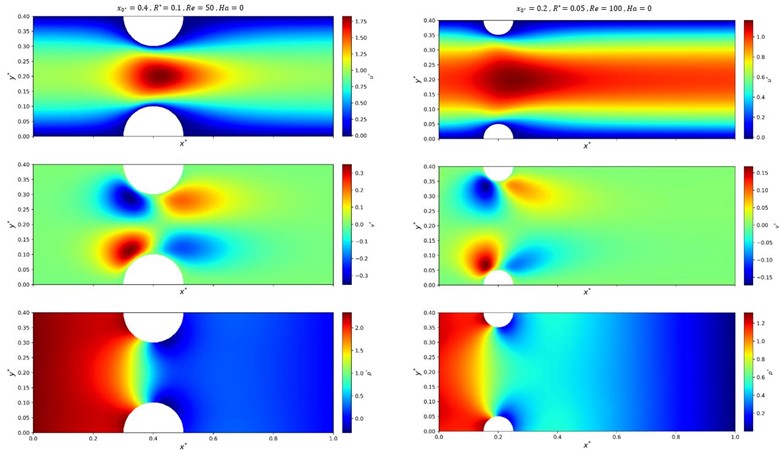}
		\caption{Pressure and velocity component contours derived from parametric solution obtained from PINN}
		\label{Fig5}
	\end{figure}

\begin{figure}[h]
		\centering
		\includegraphics[scale=1]{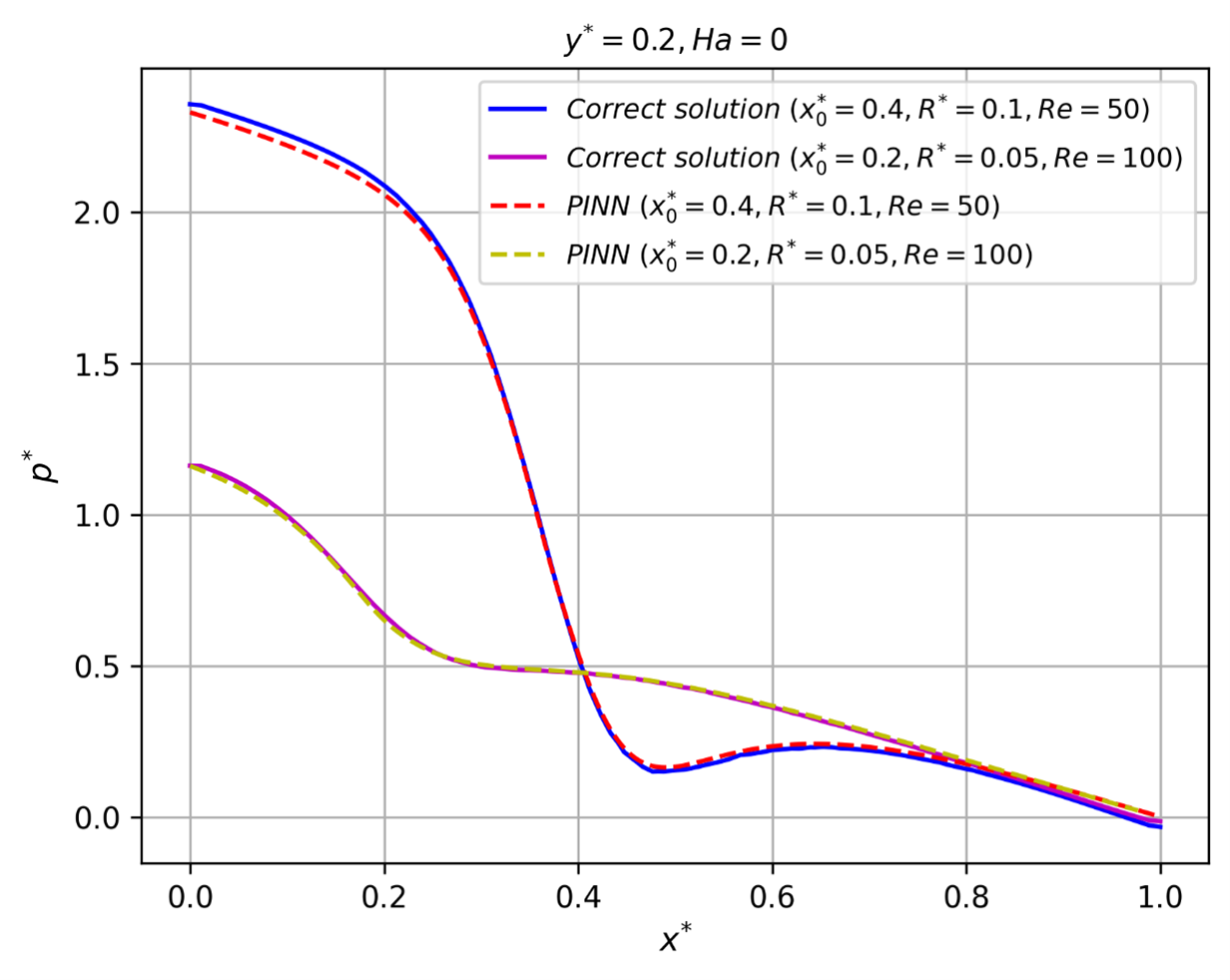}
		\caption{Trend of dimensionless pressure along the centerline of the channel and comparison with correct results.}
		\label{Fig6}
	\end{figure}

\begin{figure}[h]
		\centering
		\includegraphics[scale=1]{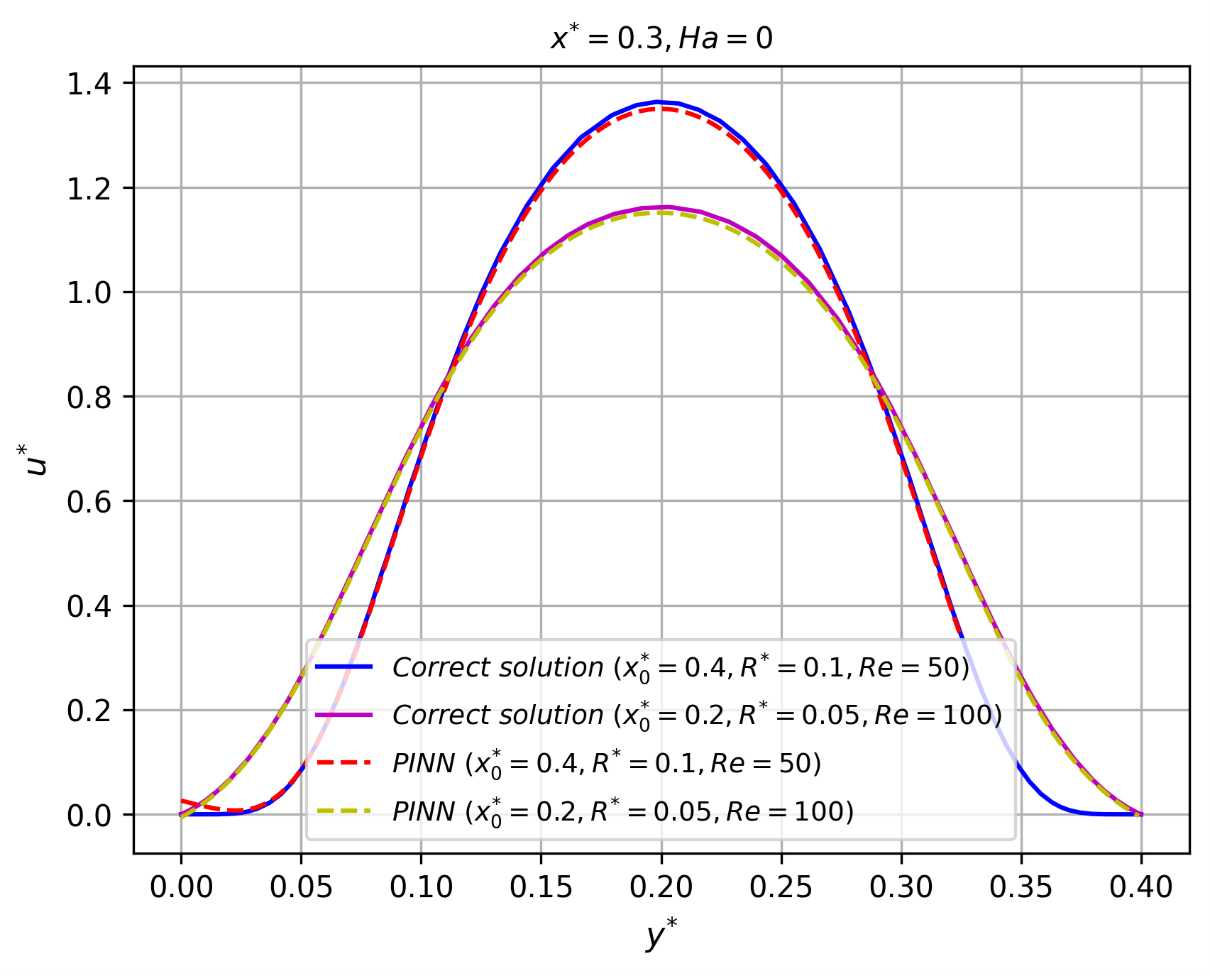}
		\caption{Comparison of the velocity component in x direction in a section of the channel obtained by PINN and the correct results.}
		\label{Fig7}
	\end{figure}

	In this section, use of PINN‌ in analyzing the flow inside an obstructed channel as described in \ref{Fig2} is investigated. Range of physical dimensionless parameters ($Ha,Re$) and geometric dimensionless groups($R^\ast, x_0^\ast$) is described in the following Table:
	\begin{table}[h]
		\centering
		\begin{tabular}{|c|c|}
			\hline
			Re&20-120\\
			\hline
			Ha&0-22\\
			\hline
			$R^\ast$&0.02-0.12\\
			\hline
			$x_0^\ast$&0.15-0.5\\
			\hline
		\end{tabular}
		\caption{Dimensionless group ranges}
	\end{table}
	Since the obtained solution is parametric, physical and geometric parameters can be studied in the desired range. To visualize the results, two random numbers in the parameter ranges are selected and their dimensionless velocity component contours as well as pressure is shown in Fig.\ref{Fig5}
	According to Fig.\ref{Fig5}, in some regions the pressure is negative which result in vortices inside the channel. In the region after the obstruction, instabilities and flow separation can be caused by positive pressure gradient. At the separation point, the force resulted from pressure gradient overpowers the inertia of the fluid, causing the flow to separate from boundary layer. Furthermore, at the regions before the obstruction, formation of dead areas is probable. In order to verify the obtained results, a conventional method has been employed. As depicted in Fig\ref{Fig6}, pressure along the centerline follows a similar trend similar to the one obtained from PINN. 
	\FloatBarrier
	\FloatBarrier
	Moreover, by examining velocity profile along the section of channel where $x^\ast=0.3$ a well-established conformity is perceived according to Fig.\ref{Fig7}. An increase in maximum velocity is seen at the narrow part of the tube as a result of mass conservation. Similarity of the results obtained from PINN and other computational methods proves it's robustness.
	\FloatBarrier
	\FloatBarrier
	\subsection{Analyzing the flow when magnetic fields are applied}
\begin{figure}[h]
		\centering
		\includegraphics[scale=1]{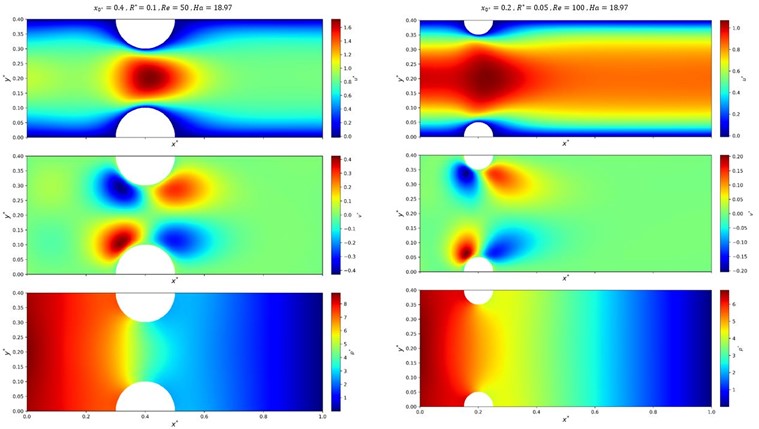}
		\caption{Contours of dimensionless velocity components and pressure when magnetic fields are applied, obtained from PINN}
		\label{Fig8}
	\end{figure}

\begin{figure}[h]
		\centering
		\includegraphics[scale=1]{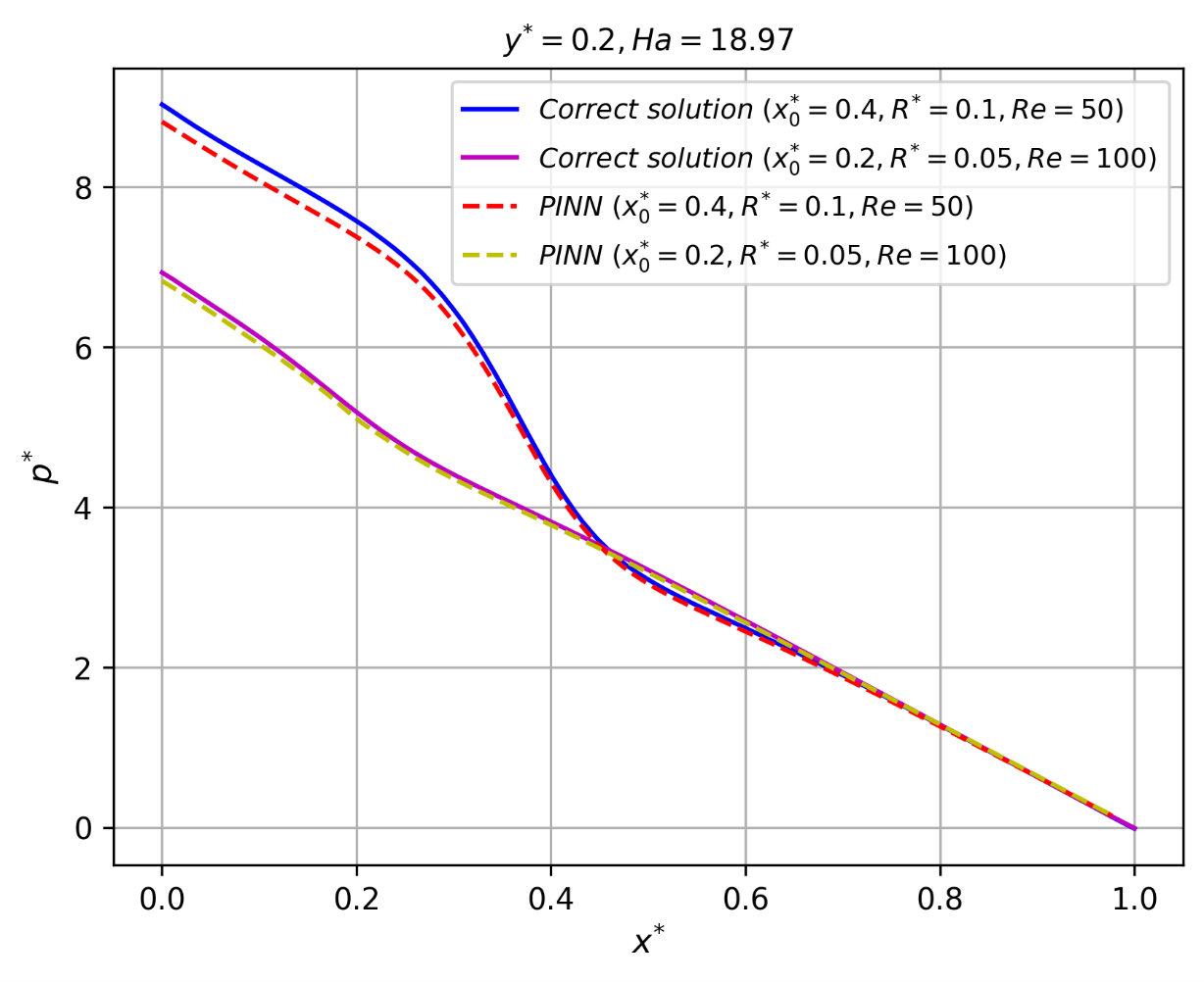}
		\caption{Comparing dimensionless pressure changes along the centerline of the channel with correct results of flow influenced by magnetic fields}
		\label{Fig9}
	\end{figure}

\begin{figure}[h]
		\centering
		\includegraphics[scale=1]{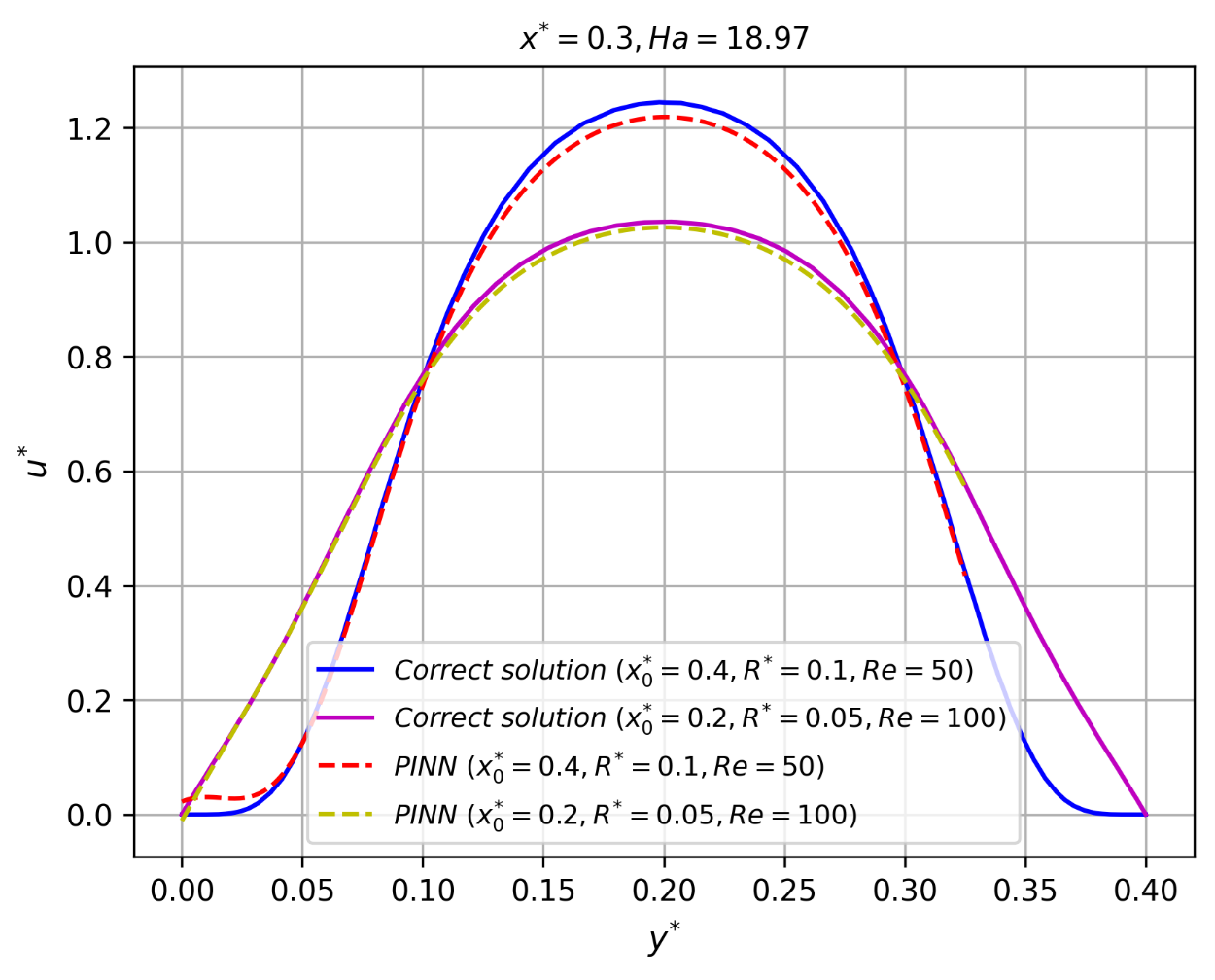}
		\caption{changes of the dimensionless velocity component in x direction in a section of the channel and comparison with the correct results for the flow under the influence of magnetic field}
		\label{Fig10}
	\end{figure}

	As the magnetic fields are applied to fluids like blood which are conductive, a source term has to be considered in the momentum equations. Effect of this term in parameters such as dimensionless pressure and velocity can be seen in Fig.\ref{Fig8} where these parameters are depicted in form of contours. According to the obtained results, as the magnetic field is applied, less vortex inducting areas are created and the separation point of the flow moves further down stream. It is evident that magnetic fields can be used to prevent creating vortices, dead areas and delays separation.
	\FloatBarrier
	\FloatBarrier
	In order to investigate the matter at hand further, dimensionless pressure along the channel centerline is depicted in Fig.\ref{Fig9}. As the magnetic field is applied, pressure loss in the channel in increased, which is useful to determine if the channel is congested or not. This pressure drop can easily detect blockage in minuscule veins, which is the initial stage of many medical complications. Applying the magnetic field causes the regions of pressure drop to shrink, preventing flow separation from happening.
	In the following section, the velocity profile of the obstructed channel under the influence of magnetic field is compared with data from other sources and CFD. By studying the velocity profile in different sections of the channel as shown in Fig.\ref{Fig10}, it can be deducted that appliance of the magnetic field smoothens the velocity profile, making it like the turbulent flow, which can delay separation in the channel.
		\FloatBarrier
	\FloatBarrier
	\FloatBarrier
	\FloatBarrier
	\subsection{Extrapolation beyond the range of parameters}
\begin{figure}[h]
		\centering
		\includegraphics[scale=1]{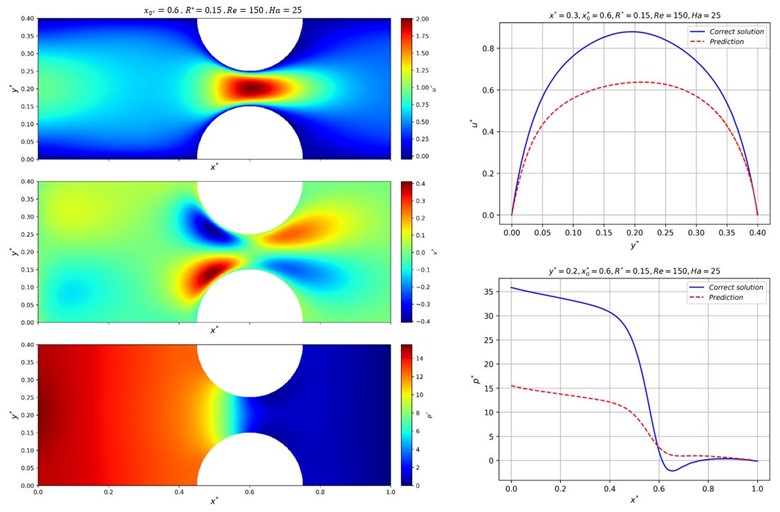}
		\caption{Comparing results calculated by PINN with the correct results}
		\label{Fig11}
	\end{figure}

	In order to assess the capabilities of the described method for analyzing the flow outside of designated ranges, some out of range parameters as described in Fig\ref{Fig11} are provided as input. The obtained dimensionless pressure and velocity fields deviated from the correct results, yet the precision of the results was acceptable. Indicating the ability of this model to handle unseen cases. The dimensionless pressure observed along the centerline of the channel deviates from the results obtained from CFD models, implying that the results are not accurate which is expected considering the fact that all of the input parameters are out of range. 
	\FloatBarrier
	\FloatBarrier
	\subsection{Extrapolating the problem for unobstructed case(Poiseuille flow solution)}
\begin{figure}[h]
		\centering
		\includegraphics[scale=1]{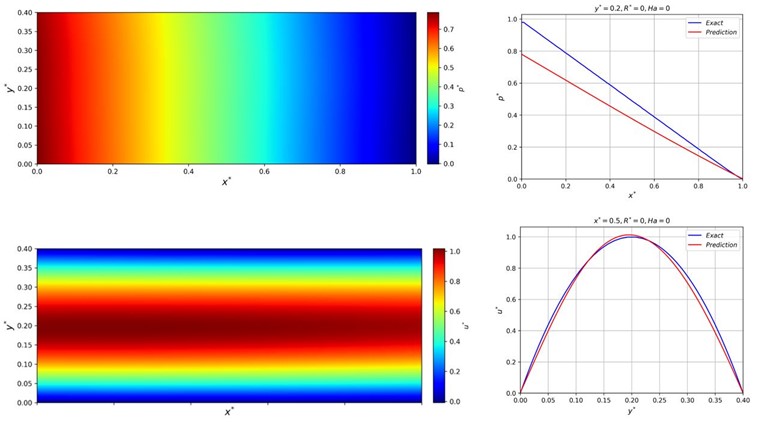}
		\caption{Comparison of the results obtained from PINN‌ with analytic solution of the Poiseuille flow}
		\label{Fig12}
	\end{figure}

	In this section, the flow inside of the unobstructed channel has been parametrically extrapolated using PINN and compared with analytical solution of the Poiseuille flow. The dimensionless velocity at the center of the channel as well as pressure drop along the centerline connecting the inlet and outlet of the channel has been compared to the results obtained from analytical solutions. According to Fig.\ref{Fig12}, the results obtained from PINN slightly differ from the results obtained from the analytical solutions, despite the significantly different geometry of the domain.
	This indicated the ability of this method to be generalized in order to solve different cases of the problem.
	\FloatBarrier
	\FloatBarrier
	\subsection{The inverse problem(determining physical properties of the problem)}
\begin{figure}[h]
		\centering
		\includegraphics[scale=1]{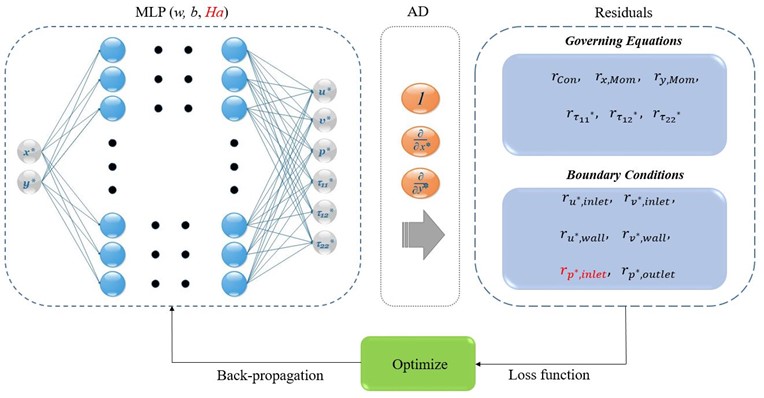}
		\caption{Schematics of the physics informed neural network used to solve the inverse problem}
		\label{Fig13}
	\end{figure}

	In the last section of the article at hand, the inverse of the described problem has been explored using PINN‌ with the geometry and sampling point distribution shown in Fig.\ref{Fig3}. Solving these types of problems using conventional methods is often challenging, yet the flexibility of PINN‌ method enables it to handle such ill-posed 
	problems with ease. For instance, here we have assumed the Hartmann number, which is a very important parameter for this physic, to be unknown. Due to the addition of an unknown to the problem, it is necessary to have velocity or pressure values of some points within the domain. In this study, a scenario has been considered particularly in medical applications where the pressure drop of a vein is measured and the properties of the magnetic field needed for countering excessive pressure drop are desired. Accordingly, the inlet pressure of the chanel is defined as a new boundary condition in addition to Eq.\ref{Eq31}
	\begin{equation}
		r_{p^\ast,inlet}=p^\ast\mathrm{\mathrm{-}}\frac{p_{\mathrm{inlet}}}{\mathrm{\rho}u_{max}^2}
		\label{Eq31}
	\end{equation}
	Additionally, Eq.\ref{Eq29} has been modified to accommodate the new problem:
	\begin{multline}
		{Loss}_{BC}=\frac{1}{N_{BC,inlet}}\sum_{j=1}^{N_{BC,inlet}}{\left({r_{u^\ast,inlet}}^2+{r_{v^\ast,inlet}}^2+{r_{p^\ast,inlet}}^2\right)}+\\
		{\frac{1}{N_{BC,wall}}\sum_{j=1}^{N_{BC,wall}}{\left({r_{u^\ast,wall}}^2+{r_{v^\ast,wall}}^2\right)+\frac{1}{N_{BC,outlet}}\sum_{j=1}^{N_{BC,outlet}}\left({r_{p^\ast,outlet}}^2\right)}}
		\label{Eq32}
	\end{multline}
	In other words, to obtain the desired variable, the inlet pressure of the channel has been provided as a new loss function(Fig.\ref{Fig13}) to enhance the training process of Hartmann number. Now the Hartmann number has to be optimized in the training process like other parameters such as network parameters including weights and biases. 
	\FloatBarrier
	\FloatBarrier
	\begin{figure}[h]
		\centering
		\includegraphics[scale=1]{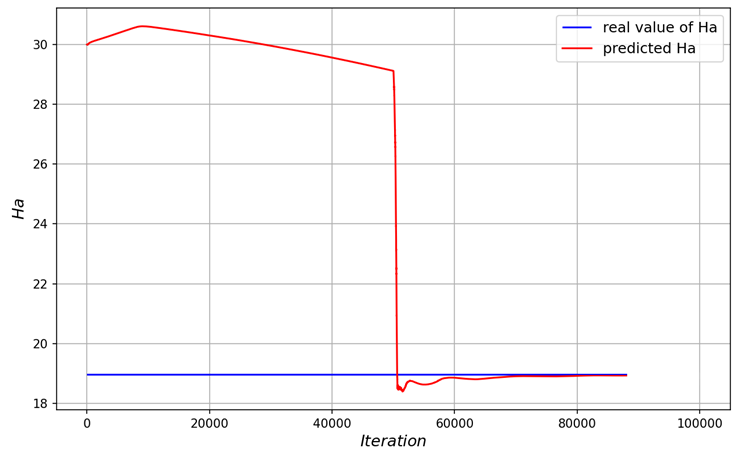}
		\caption{Calculated  Hartmann number as the training process progressed}
		\label{Fig14}
	\end{figure}
	\FloatBarrier
	As the weights and biases of the network were initialized, the  Hartmann problem has been assigned a random value. if this value is too deviant from the correct value, the solution process becomes harder, requiring more iterations to achieve the desired accuracy. In Fig.\ref{14}, the process of obtaining Hartmann number using PINN is shown. The results of the optimization converged after iteration 86000, leading to a  Hartmann number of 18.97. 
	The results of the present study show that this method is capable of predicting the flow inside of an obstructed channel under the influence of magnetic field. One of the pronounced advantages of this method is not needing labeled data for the network input variables. The process of generating data suitable for learning networks is difficult and time consuming, not even possible in some instances.\cite{31} The described method in this article has high ability to solve engineering inverse and parametric problems compared to other methods available. Combining this method with other calculation methods might result in enhanced accuracy of the results, as discussed in \cite{32}. More studies regarding network hyperparameters and structures in more complex problems can be carried out.\cite{33} The method presented in this study is applicable to more complex cases such as cases with complex geometry or cases where the fluid properties are not constant.
	\section{Conclusion}
	In this study, fluid flow inside the obstructed channel under the influence of magnetic field has been analyzed using the PINN method. Governing equations have been rewritten to incorporate derivatives of lower order.Also the governing equations have been written in dimensionless form. Dimensionless groups and geometry of the problem were provided as the input to the network. The placement and radius of the obstruction as dimensionless parameters, Hartmann number and Reynolds number have been included in the learning process and the problem has been solved parametrically. Results show good conformity with the correct data inside the predefined range and have had acceptable predictions regarding the general trend of parameters outside of the predetermined range. Inverse problem of this case has also been explored in this article, mainly in form of determining the Hartmann number using the described method. Solution of the inverse problem has also been proven to conform well to the correct results, indicating it's high accuracy. In this study, a thorough research regarding physical and geometric parameters on the case of flow inside an obstruction channel under the influence of magnetic field has been carried out by solving dimensionless and low-order form of governing equations using PINN.

	\section*{CRediT authorship contribution statement}
	\textbf{Ehsan Ghaderi}: Conceptualization, Methodology, Visualization, Software, Writing the original draft.\\
	\textbf{Bijarchi et al.}: Conceptualization, Review and Editing, Supervision.
	\section*{Acknowledgment}
	We also like to thank Nvidia for providing the GPUs for this work. Thanks should also go to Mohammadreza Sayah and Atta Shojaee Baghini.
	\section*{Conflict of Interest}
	The authors declare no conflict of interest
	\section*{Data Availability}
	The python codes used to generate the results presented are available on Github.\\
	\url{https://github.com/ehsangh94/PINN}

\end{document}